\documentclass[11pt,twocolumn]{article}
\usepackage{amsmath}
\usepackage{amsfonts}
\usepackage{amssymb}
\usepackage{authblk}

\usepackage{abstract}  
\usepackage{graphicx}
\usepackage{hyperref}
\hypersetup{colorlinks=True,linkcolor=blue,citecolor=blue,urlcolor=blue}
\usepackage[square,sort&compress,comma,numbers]{natbib}
\usepackage{fancyhdr}

\title{\textbf{ Analysis of doorway states in a graphene structure}}

\author[1]{\textbf{E. A. Carrillo\thanks{eCarrillo@fisica.unam.mx}}}
\affil[1]{Instituto de F\'isica, Universidad Nacional Aut\'onoma de M\'exico - P.O. Box 20-364, 01000 Mexico City, Mexico}
\author[1]{\textbf{ J. Flores}}

\author[1]{\textbf{G. Monsivais}}

\date{\today}

\begin{document}
\twocolumn[
  \begin{@twocolumnfalse}
\maketitle

\begin{abstract}
Doorway states, which are related to the strength function phenomenon and giant resonances, arise when two systems interact, one with a high density eigenvalue spectrum and the other with a comparatively low density. These concepts, first studied in nuclear physics in the 40's, are here analyzed from a theoretical point of view in special and simple graphene structures, obtained after applying appropriate voltages to a graphene sheet. The influence of the doorway states on the electronic transport in these systems  is also studied. To analyze these effects we consider a two-dimensional model of two potential barriers of equal height but very different widths separated by a well.
\rule{0.92\textwidth}{0.8pt}

\end{abstract}
\end{@twocolumnfalse}
]
\saythanks

\section*{Introduction}

Each new important discovery in physics has brought with it a new vision of the concepts and phenomena of nature. This occurred in solid state physics with the discovery of the first two-dimensional material, that is, graphene, which was obtained in 2004 by mechanical exfoliation of graphite \cite {Nov1, Nov2, Zha}. Since then, researchers from different disciplines of science and technology \cite {Sch, Gei, Hil, Ohi} have turned their attention to this material due to the extraordinary properties it exhibits.

The graphene is made up of carbon atoms arranged in a honeycomb-shaped sheet, with thickness of one carbon atom ($ 0.1 \mathrm {nm} $). It has a very particular band structure, as it is a semi-metal with a zero energy gap, and a linear dispersion relation with conical shape (the well-known Dirac-cones) near the points $ K $ and $K'$ (Dirac points) in the Brillouin zone \cite{Wal}. It has been shown \cite{Nov1} that electrons behave then as relativistic particles even though they move at a speed $ v_F \approx c / 300 $, where c is the speed of light, so $v_F << c$. 
\begin{figure*}[t]
\includegraphics[width=\textwidth]{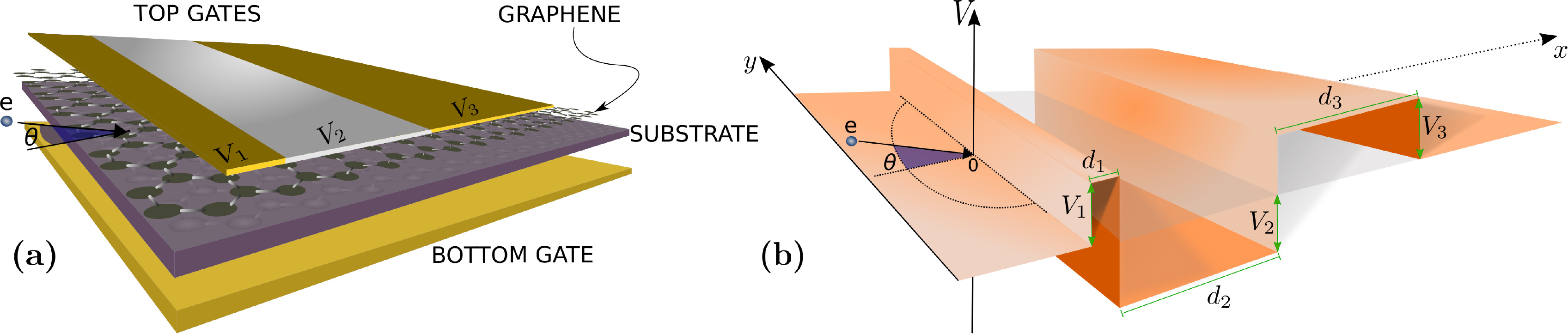} 
 
\caption{\label{fig:epsart} \textbf{(a)} Schematic diagram of the device. The graphene sheet is placed on a substrate which does not modify the electronic properties. Three electrodes are deposited over the graphene sheet. They are kept at voltages $V_1$, $V_2$, and $V_3$. Figure (b) shows the two-dimensional potential profile produced by the electrodes in \textbf{(a)}, where the height of the different regions of the potential are equal to the voltages $V_1$, $V_2$, and $V_3$, respectively.  The potential has two barriers and a well. The angle of incidence $\theta$ is on the plane of the graphene sheet.}
\end{figure*}
A consequence of this particular characteristic is that the probability for electrons to pass through a potential barrier created on a graphene layer is equal to one when they impinge along the layer normally to the barrier, \textit{i.e.} the barrier will be transparent for $\theta = 0$ in figure \ref{fig:epsart}\textbf{(b)}. This effect, known as Klein tunneling \cite{Kat3,STAN,Young2009}, contrasts with the case of non-relativistic electrons where the probability that a particle is transmitted decays exponentially with the height of the barrier. Klein tunneling occurs not only through a single potential barrier but also through many barriers \cite{Bar,heraclio,carrillo}, not necessarily forming a super lattice. When the electrons impinge with an angle different from zero, the Klein tunneling disappears and instead the system presents a set of resonances with transmittance equal to one. 

The purpose of this paper is to analyze properties of graphene structures which have a doorway state \cite{bohr,mor2012,KAWATA,2014doorway,2019doorway,PhysRevLett.98,PhysRevLett.105,PhysRevLett.100,Franco_Villafa_e_2011}. Therefore, in this work the intersection of two important different fields: the physics of two-dimensional materials and that of doorway states is studied. Doorway states are present in nuclei, atoms, molecules and classical systems, such as microwave resonators or sedimentary basins. In all these cases, a system with few eigenvalues called ``distinct states'' interacts with a second one with a spectrum, called the ``sea of states'', which has a much higher eigenvalue density. The distinct states are called doorway states in the literature. Thus, a composite system is formed in which the doorway states will somehow modulate the sea states. The strength function phenomenon then appears, which means that the excitation intensity of the eigenestates of the composite system is modulated by the doorway states, giving rise to an enveloping curve with a Lorentzian-like shape \cite{bohr}.

To design the graphene system having doorway states we must proceed in a different way than in non-relativistic quantum mechanics since the electrons in graphene behave quasi-relativistically, so they obey Dirac equation instead of Schrödinger equation. In the non-relativistic case, the potential that produces doorway states consists of a narrow well coupled to a much wider well, while in graphene we find that the potential that produces doorway states is formed by a thin barrier coupled to a wider barrier. 
 
\section*{Model and Method}
The composite system we analyze in this work is generated by placing metallic contacts or gates kept at voltages $V_1,V_2$ and $V_3$, respectively, on a graphene sheet deposited on a non-interacting substrate, such as $ \mathrm{SiO_2} $; see figure \ref{fig:epsart}(a). The corresponding potential has two barriers and a well, as shown in figure \ref{fig:epsart}(b). The first barrier, which will be called the thin barrier, has height $V_1 $ and width $d_1$. 
In-between the two barriers there is a well with depth $V_2 = -V_1$ and width $d_2$. The second barrier is wider than the first one so $d_1 << d_3$ but $d_3 = d_2 $ and $V_3 = V_1$.  The thin barrier generates the doorway states.

When the barriers are generated by means of electrostatic fields, changes in the Dirac cones occur only in the regions where the electrodes are localized, as seen in figure 1 of reference \cite{NovoselovNature}. The particle moves on the XY plane and the transmission largely depends on the width of the barriers and the angle of incidence.

The transmission properties of the composite system can be computed using the transfer matrix method \cite{PYEH, PMAR}. To apply this methodology we need the dispersion relations, the wave vectors, and the wave functions in the barriers, in the well and in the left and right semi-infinite regions. In the semi-infinite regions the dispersion relations are 

\begin{equation}
E = \pm \hbar v_F k,
\label{eq:1}
\end{equation}

\noindent and the wave functions are
\begin{equation}
\psi_{k}^{\pm} = \frac{1}{\sqrt{2}}
\left(\begin{array}{c}
1 \\
u_{\pm}
\end{array}\right)
e^{\pm ik_{x}x + ik_{y}y},
\end{equation}

\noindent where $ v_F $ is the Fermi velocity, $ k $ is the magnitude of the wave vector $\bf k$ at the semi-infinite regions, $ k_x $ and $ k_y $ are the longitudinal and transversal components of $\bf k$, and $ u _{\pm} = \mathrm{sgn}(E) e^{\pm i \theta} $ are the coefficients of the wave functions that depend on the angle of incidence of the electrons, $ \theta = \mathrm{arctan}(k_y /k_x) $.

The kinetic energies in the region of the well and the barriers are 

\begin{equation}
 E-V_i = \pm \hbar v_F q, 
 \label{eq:2}
\end{equation}

\noindent and the eigenfunctions are

\begin{equation}
\psi_{q}^{\pm} = \frac{1}{\sqrt{2}}
\left(\begin{array}{c}
1 \\
v_{\pm}
\end{array}\right)
e^{\pm iq_{x}x + iq_{y}y}.
\end{equation}

\noindent Here $q$ is the magnitude of the wave vector $\bf q$ in these regions, $q_x$ and $q_y$ are the components of $\bf q$, and $v_{\pm}$ are the coefficients of the wave
functions.

By imposing continuity of the wave function and conservation of the transverse momentum $q_y=k_y$ at the boundaries, we can obtain the transfer matrix \textbf{M} of the structure \cite{PYEH,PMAR}

\begin{equation}
\bf{M} = M_{I_L} M_{B_1} M_W M_{B_2} M_{I_R},
\end{equation}

\noindent with $\bf{M}_{I_L}$ and $\bf{M}_{I_R}$ the transfer matrices associated with the left and right semi-infinite regions, $\bf{M}_W$ with the well and $\bf{M}_{B_1}$ and $\bf{M}_{B_2}$ with the thin and wide barrier, respectively. These matrices depend on the so-called dynamic and propagation matrices $\bf{D_i}$ and $\bf{P_i}$ \cite{Rodriguez2012,GARCIACERVANTES2015}:

$$
\bf{M}_{I_L} = (D_{I_L})^{-1} P_{I_L} D_{I_L}
$$

$$
\bf{M}_{B_1} = (D_{B_1})^{-1} P_{B_1} D_{B_1}
$$

$$
\bf{M}_W = (D_W)^{-1} P_W D_W
$$

$$
\bf{M}_{B_2} = (D_{B_2})^{-1} P_{B_2} D_{B_2}
$$

\begin{equation} 
\bf{M}_{I_R} = (D_{I_R})^{-1} P_{I_R} D_{I_R}.
\end{equation}

The transmission coefficient of the Dirac electrons is given by the (1,1) element of the transfer matrix $\bf{M}$ \cite{PYEH, PMAR}:
\begin{equation}
  { \left\vert t(E,\theta)\right\vert}^2 =\frac{1}{\vert M_{11} \vert^2}.
\end{equation}

\section*{Results and Discussion}

\begin{figure}[!ht]
\centering
\includegraphics[width=.49\textwidth]{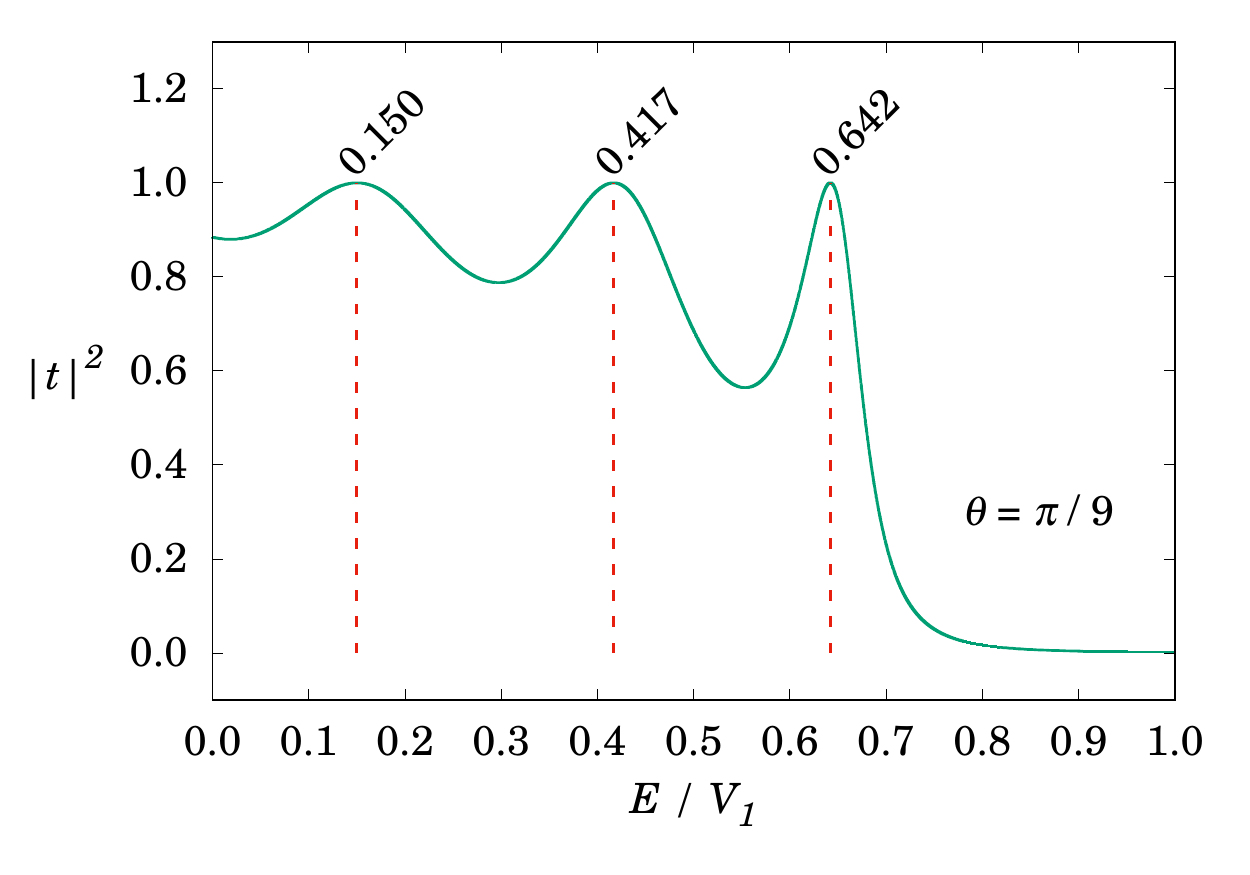} 
\includegraphics[width=.49\textwidth]{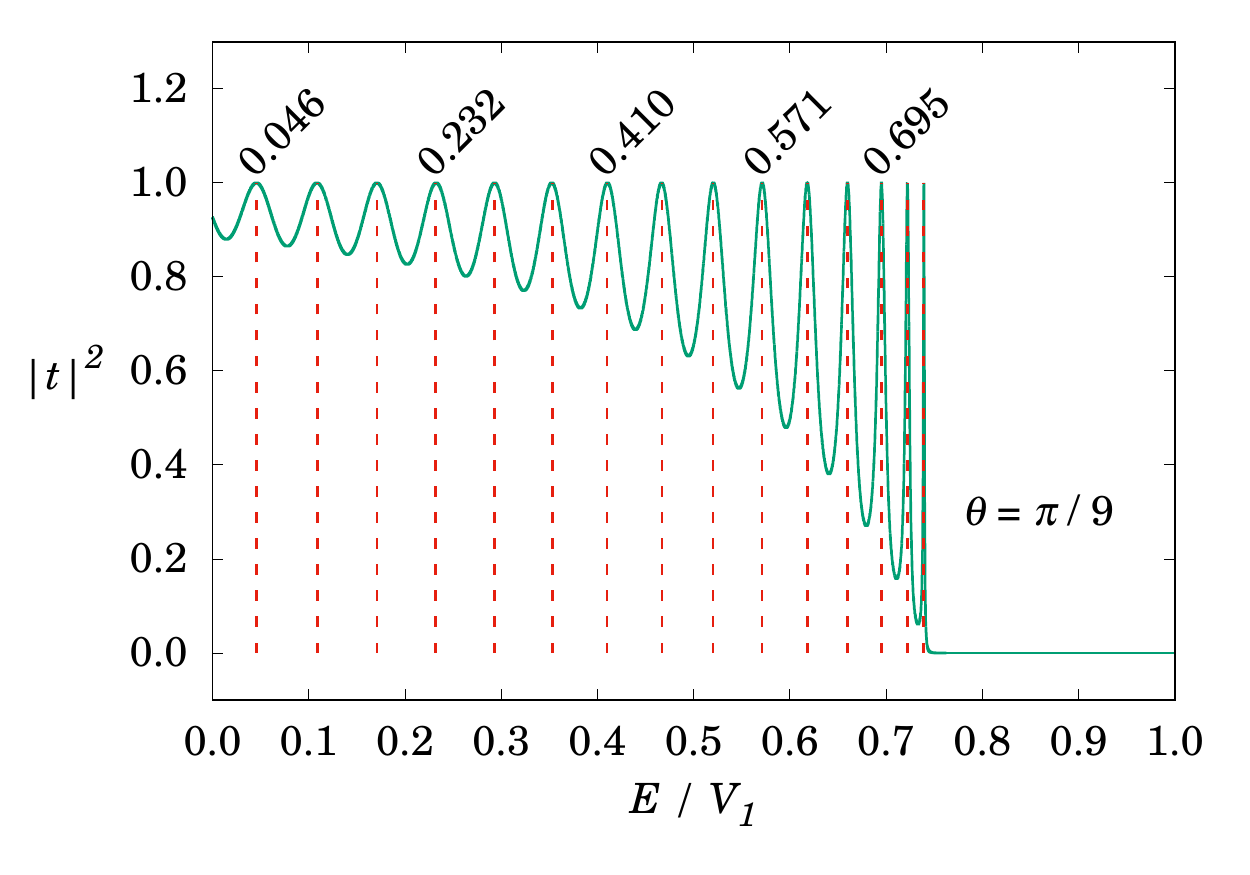} 
\caption{\label{fig:res1} Transmission coefficient $\vert t \vert^2$ as a function of $E/V_1$ for a single square barrier equal to the left one in figure \ref{fig:epsart}(a). Here $V_1=1$eV and $\theta = \pi / 9$. In (a) $d_1=45a$ and in (b) $d_1=200a$.
}
\end{figure}

We first study a single barrier and calculate its transmission coefficient. We focus on a range of energies below the barrier potential. In figure \ref{fig:res1}(a) $d_1=45a$  and in (b) $d_1=200a$, where $a$ is the carbon-carbon distance in graphene, which is equal to $0.142$ nm. In figure \ref{fig:res1}(a) the transmission coefficient shows three maxima whereas in (b) it shows fifteen. At these maxima, whose positions are indicated with dashed vertical lines, $\vert t\vert^2=1$. 
Using equations \ref{eq:1} and \ref{eq:2}, and the necessary condition for a resonance $q_x d_1 = n \pi $ \cite{Kat3},  we obtain an analytical expression for the number of maxima in these graphs.

The energy must fulfill

$$
q_x d_1 \; =d_1\sqrt{\left(\frac{E-V_1}{\hbar v_F}\right)^2-k_y^2}  
$$

$$
\qquad =d_1\frac{V_1}{\hbar v_f}\sqrt{\left(\frac{E}{V_1}-1\right)^2-\left(\frac{E}{V_1}\sin\theta\right)^2}
$$

\begin{equation}
\qquad = n \pi,
\end{equation}

\noindent where $\hbar v_f \approx 4a\:\mathrm{eV} = 4 \times0.142 \:\mathrm{eV} \cdot \mathrm{nm} $. From this equation we can see that there is a finite number of values of $n$ for which $0<E/V_1<1$. For example, with $\theta=\pi/9$ and $d_1=45a$ one has $n=1,2,3$, whereas changing $d_1$ to $200a$ one obtains $n=1,\ldots,15$. This matches the results shown in figures \ref{fig:res1}(a) and \ref{fig:res1}(b). It should be noted that in both figures the maxima are equal to 1, the only difference being the density of resonances, which grows with the barrier width.

\begin{figure}[!ht]
\includegraphics[width=.49\textwidth]{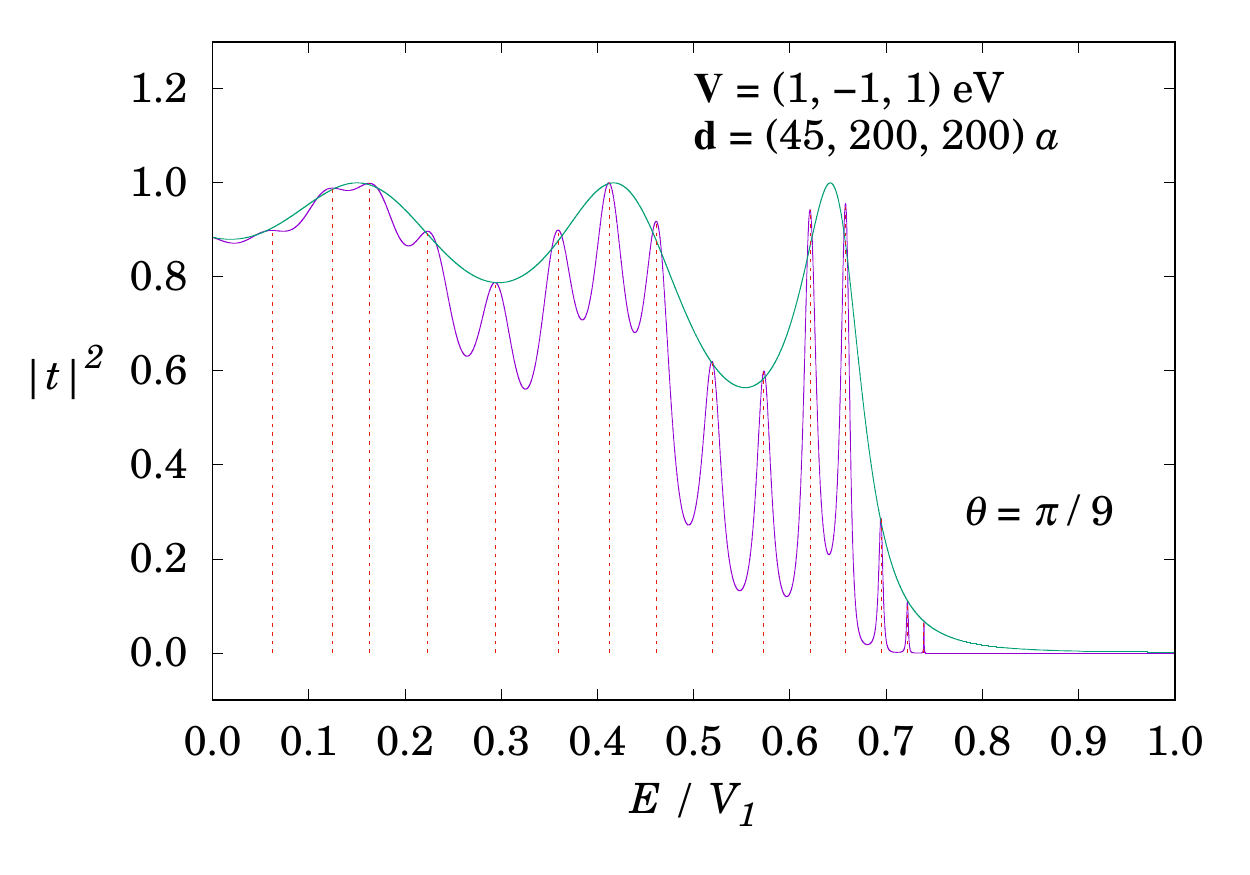} 
\includegraphics[width=.49\textwidth]{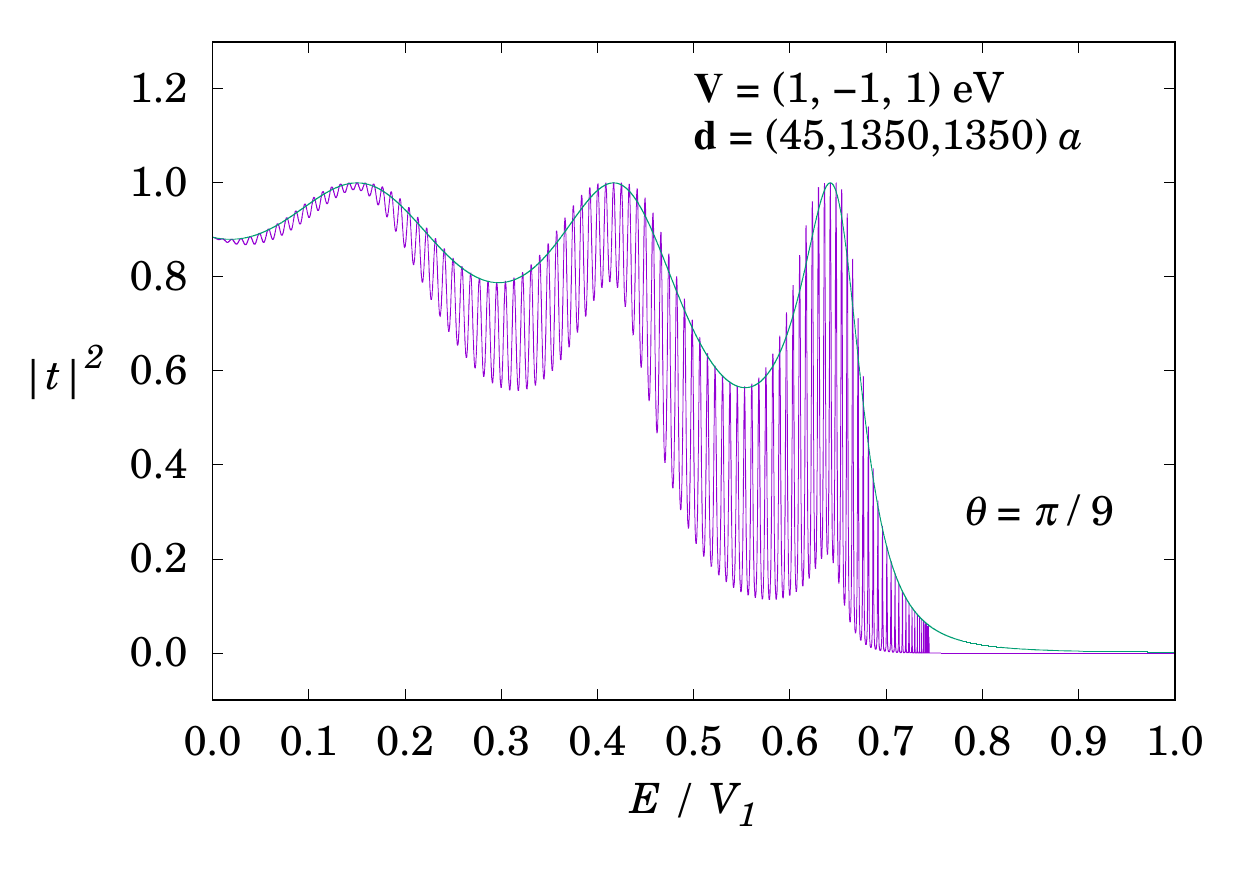}
\caption{\label{fig:res23} The transmission coefficient through a single barrier (green line) modulates the transmission through the composite system (purple line). Here ${\bf V} =(V_1,V_2,V_3)$ and ${\bf d} =(d_1,d_2,d_3)$. In case (b) the number of resonances is larger than in case (a) because the widths of the well and of the second barrier are several times wider.}

\end{figure}

We now study the composite system. Its transmission spectrum is compared in figure \ref{fig:res23}  with that corresponding to the thin barrier. In figure \ref{fig:res23}(a) the parameters of the thin barrier are as in figure \ref{fig:res1}(a), while for the well and the wide barrier $ d_3 = 200a$. We can see that the 15 resonances of figure \ref{fig:res1}(b) are now modified. Their intensities (purple line) are modulated by the transmission curve corresponding to that of the single barrier (green line). Furthermore, the vertical red dotted lines indicate that the position of the maxima along the energy axis changes slightly. In general, the maxima of the purple line are different from $1.0$, so the composite system does not have a perfect transmission. However, the purple maxima whose energies are close to the energies of the green maxima are larger than the other purple maxima. 

The shape of the transmission spectrum graph can be understood by the following argument: when the waves entering the system have a frequency close to or equal to any of the resonances, a quasi-standing wave is established along the whole system. To obtain a constructive interference pattern, the waves have to travel several times back and forth within the system. It is therefore clear that the waves will settle within the small body in a shorter period of time than it takes for the waves to settle in the composite system. So resonances with frequencies close to those of the small system will absorb energy faster than the other resonances and the peaks corresponding to these resonances are higher. For a more detailed discussion about the relation between the time that a resonance takes to be established and its strength, see the comments in reference \cite{mor2012} related to its figures 3, 4 and 5.

In figure \ref{fig:res23}(b) the  well and the wide barrier widths were increased with respect to those of figure \ref{fig:res23}(a) so the transmission spectrum shows many more resonances. The system in (b) is much larger than in (a), but the behavior of the transmission coefficient is controlled in both cases by the thin barrier, so in the two figures the green curve is the envelope of the purple curve, showing the strength function phenomenon.

\begin{figure}[t]
\centering
\includegraphics[width=0.49\textwidth]{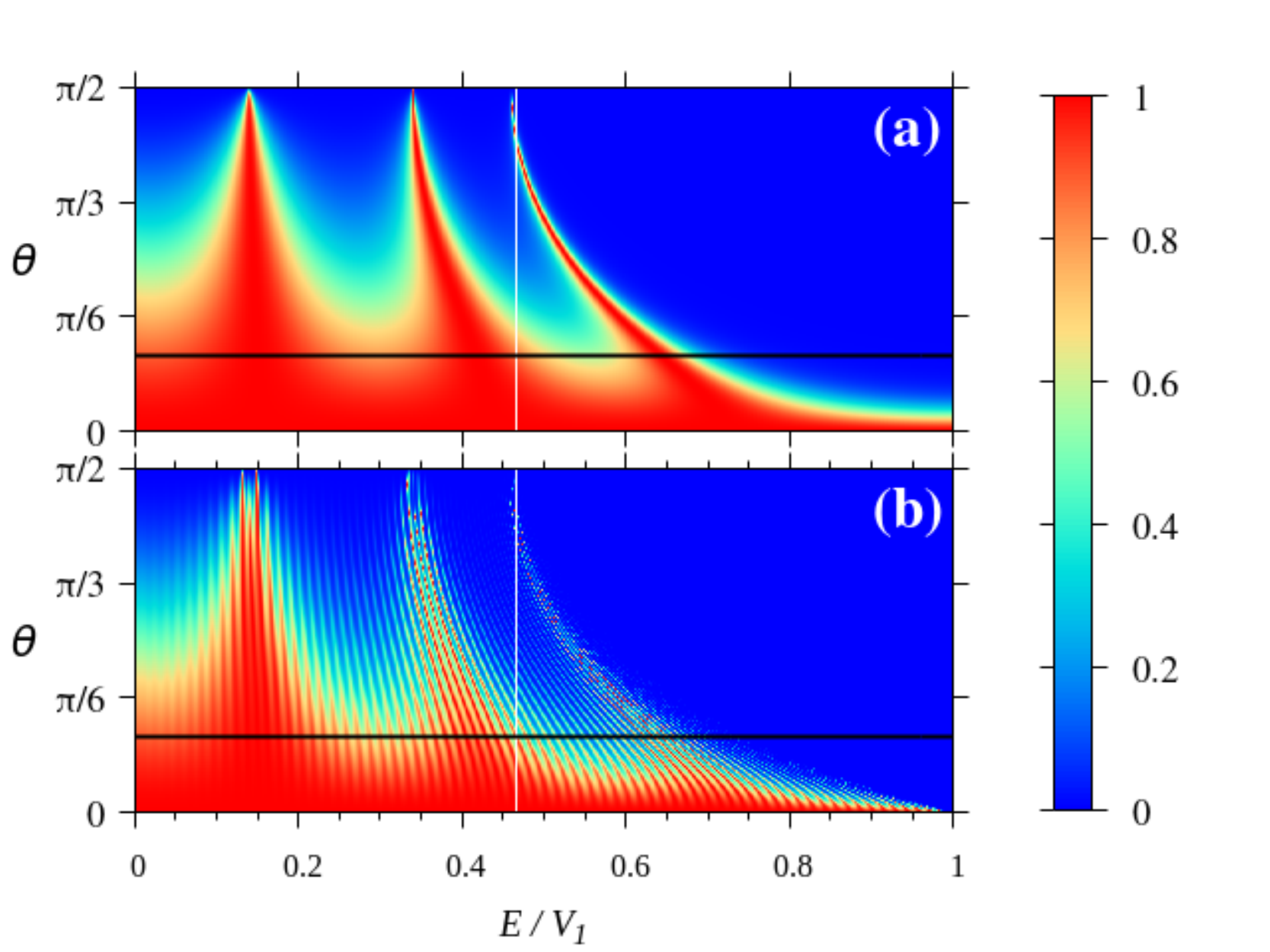} 
\caption{\label{fig:res45} Contour maps of the transmission coefficient $\vert t \vert^2$ as a function of both the normalized energy $E/V_1$ and the angle of incidence $\theta$ for one barrier (figure (a)) and for a composite system with ${\bf d} =(45,900,900)a$ (figure \textbf{(b)}). In both cases the horizontal line indicates the angle $\theta = \pi / 9$ used in figures \ref{fig:res1} and \ref{fig:res23}.
According to the Klein effect, for $\theta=0$ the transmission coefficient is equal to 1 for all energies, so throughout the abscissa all horizontal lines near the energy axis are red. The vertical white line indicates that the energy value close to the position of the third peak is the same in both figures.}
\end{figure}

Up to now, the angle at which the electrons impinge has been $\pi /9$. In what follows we shall present results for other values of $\theta$. In figure \ref{fig:res45} the contour plots are shown. Again, the transmission spectrum of the thin barrier (figure (a)) is compared with that of the composite system (figure (b)). For the thin barrier we see a spectrum with only three maxima. As the angle changes, these maxima shift slightly on the energy axis. On the other hand, for the composite system the transmission coefficient shows many maxima. However, despite the change in the angle, the maxima are still modulated by the spectrum of the thin barrier.

If we overlap figures \ref{fig:res45}(a) and  \ref{fig:res45}(b), the red zone of figure (a) covers completely the red zone of (b).
To guide the eye, the vertical white line has been inserted to indicate that the energy value $E/V_1$ of the third peak is approximately equal to $0.465$ in both figures; this is also true for the other two peaks in the graphs.

\begin{figure}[t] 
\centering
\includegraphics[width=.49\textwidth]{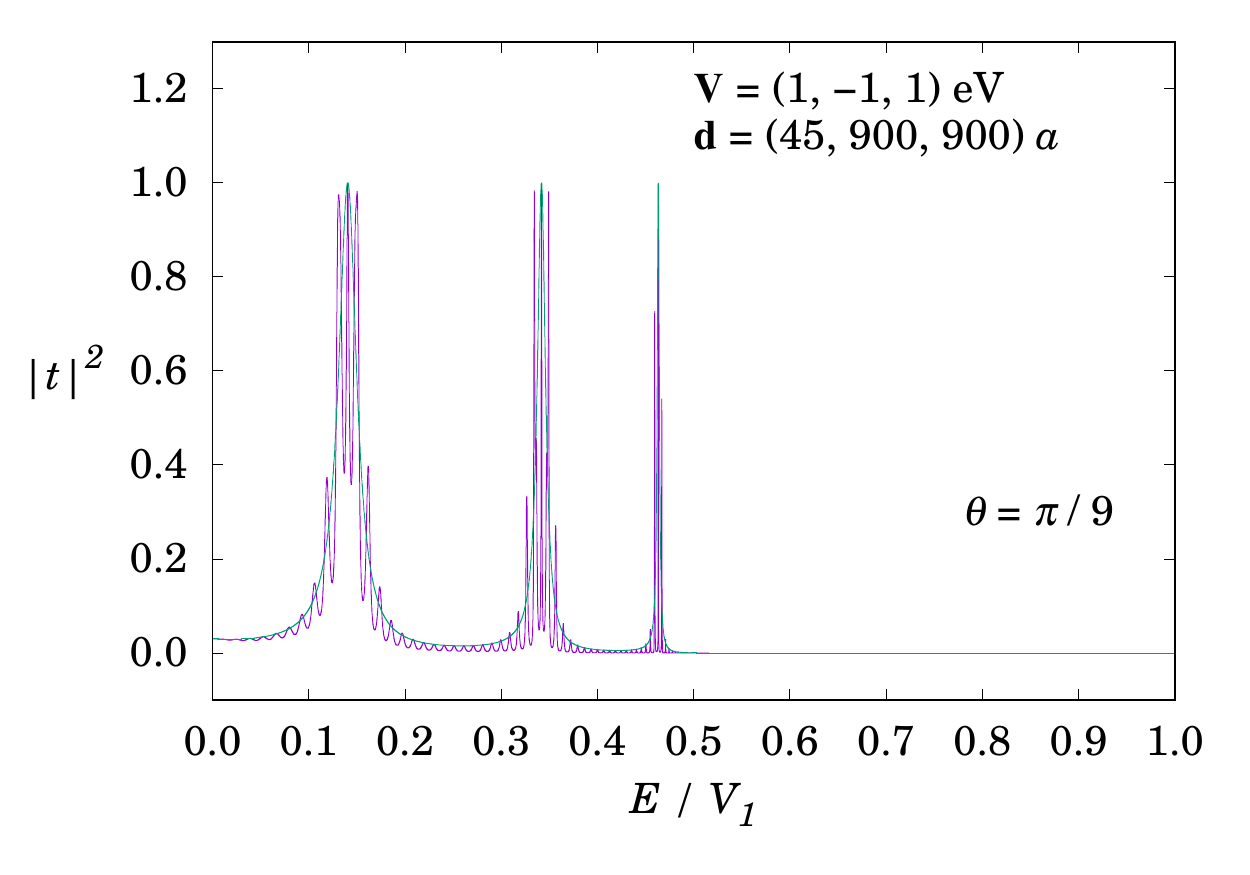} 
\caption{\label{fig:res6} Comparing this graph with the results shown in figures \ref{fig:res1} and \ref{fig:res23}, it will be seen that the width of each resonance in the transmission coefficient decreases as the angle increases.}
\end{figure}
 
We now analyze the behavior of the transmission coefficient as a function of $\theta$. In  figure \ref{fig:res6} the results for $\theta = 4 \pi / 9 $ are presented. We see that there are still three maxima and that the green curve also modulates the behavior of the purple curve. The minima  of the transmission coefficient now approach zero and the width of each resonance decreases as the angle $\theta$ increases, making the strength function phenomenon even more evident.

\begin{figure}[t]
\centering
\includegraphics[width=.49\textwidth]{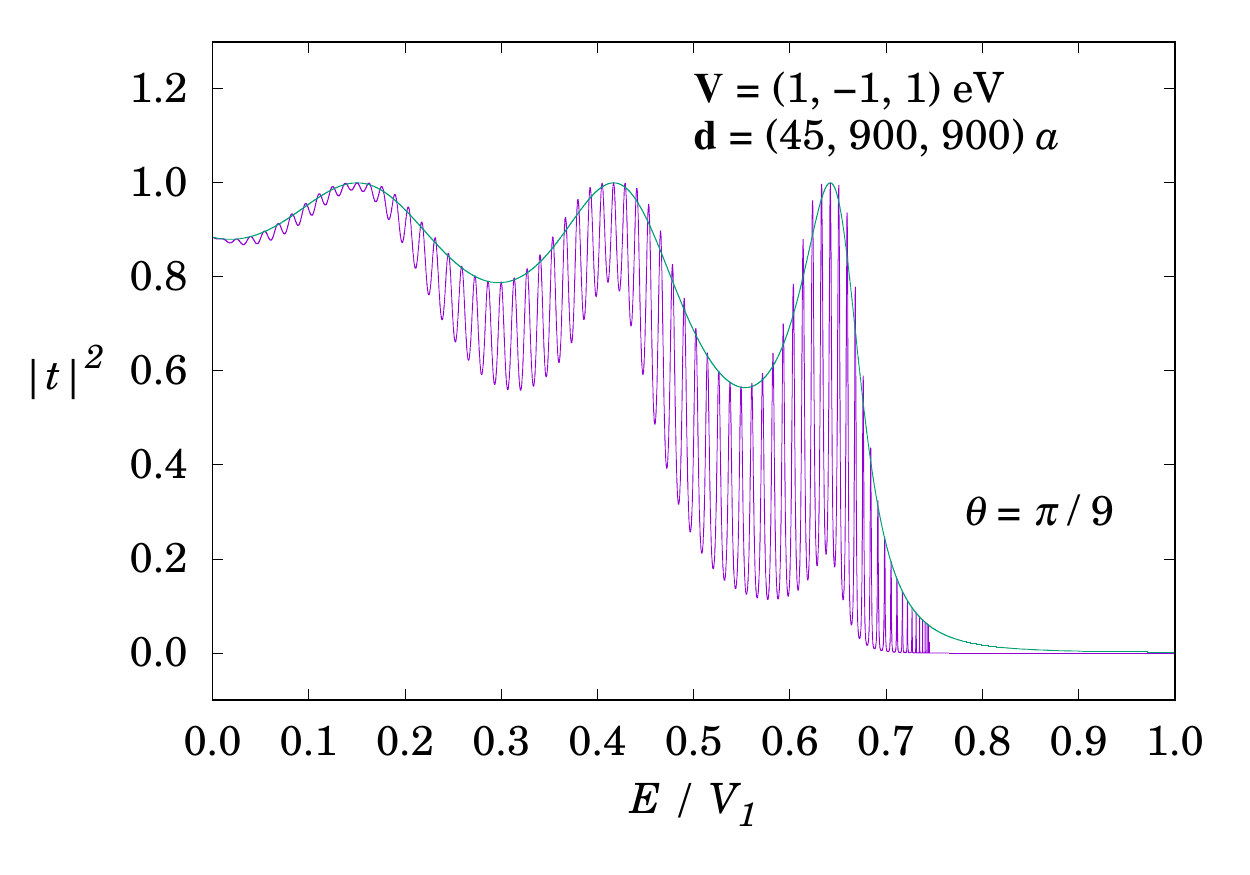}
\includegraphics[width=.49\textwidth]{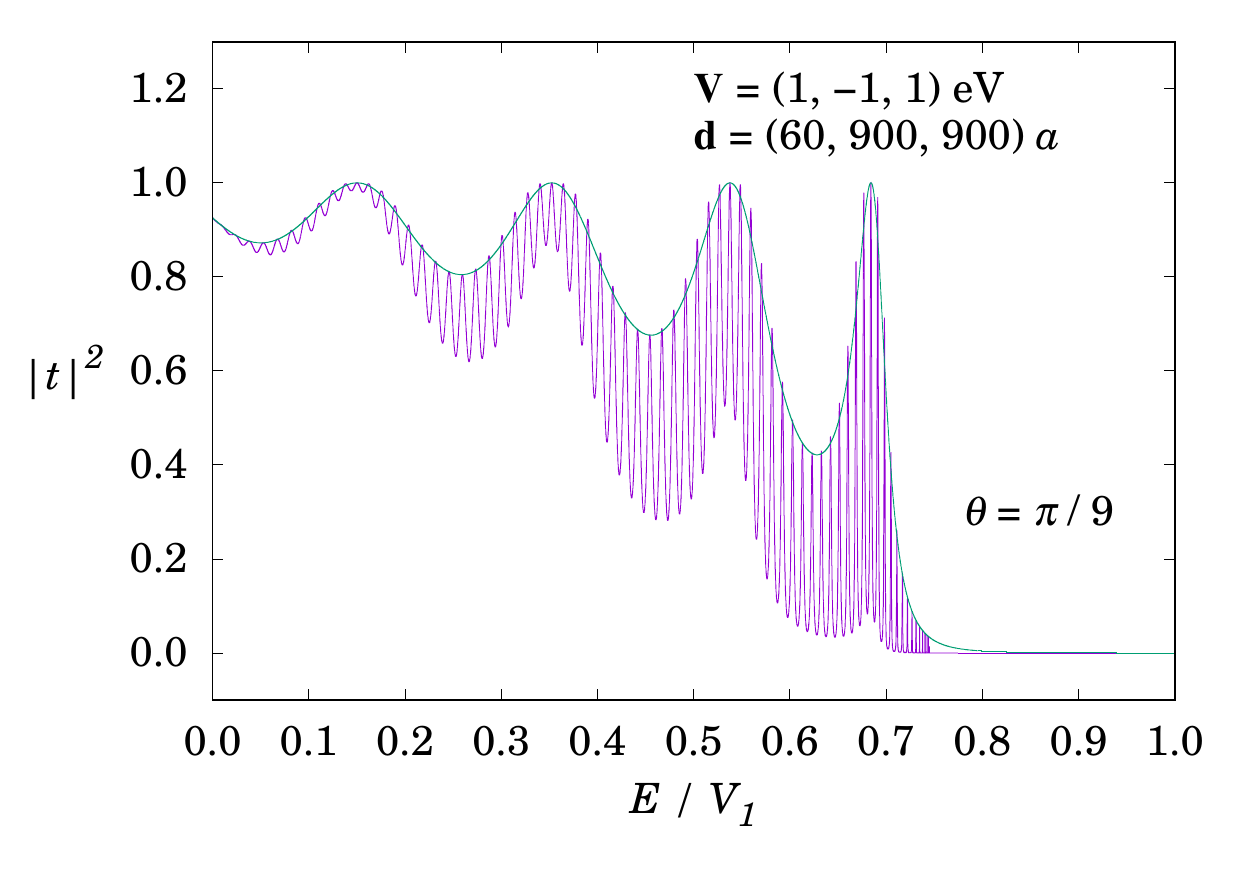} 
\caption{\label{fig:res7} The number of doorway states changes with the width of the thin barrier.}
\end{figure}

In the previous plots the thin barrier has a width equal to 45$a$. In figure \ref{fig:res7} this width is 60$a$. This increases the number of maxima, so there are now 4 peaks, but the green curve still modulates the purple one.

\section*{Conclusions}
In this work we have studied a graphene structure which has doorway states that allow a more efficient energy absorption. We analyze the electronic transport as a function of the energy and of the angle of incidence. For the thin barrier, the transmission maxima will be equal to 1.0; this occurs at the doorway state energies. Here the transmission coefficient of the thin barrier follows a smooth curve with well-localized maxima, and it turns out that this curve is an envelope for the resonances of the composite system.

In other fields of physics, when resonating systems with a doorway state are dealt with, this envelope is obtained by interpolating the individual transmission, so it is a mathematical construction \cite{mor2012}, while in the system we have studied in this paper the green curve has a physical meaning since it corresponds to the transmission for a single barrier. 

The doorway states, together with the angle of incidence variation can be useful to build optoelectronic devices, such as filters or sensors.

\section*{Acknowledgement}
E. A. Carrillo acknowledges to CONACYT-Mexico for the postdoctoral research fellowship at IFUNAM. The authors thank G. G. Naumis for his careful reading of the manuscript and his useful comments.

%\newpage
%\printbibliography%[heading=none]

\bibliography{refs}
\bibliographystyle{unsrtnat}
%\bibliography{refs}

\end{document}